\documentclass[12pt]{article}
\usepackage{times}

\usepackage{balance} 
\usepackage{amsmath}
\usepackage{amssymb}
\usepackage{float}
\usepackage{graphicx}
\usepackage{caption}
\usepackage{subcaption}
\usepackage{tabularray}

\usepackage{lineno}
\usepackage{url}
\usepackage{pdfpages}
\usepackage{setspace} 

\usepackage{soul}
\usepackage{url}
\usepackage[hidelinks]{hyperref}
\usepackage[utf8]{inputenc}
\usepackage{booktabs}
\usepackage{algorithm}
\usepackage{algorithmic}
\usepackage[comma,sort&compress]{natbib}
\usepackage{dblfloatfix}

\usepackage{draftwatermark}
\SetWatermarkScale{1.2}
\SetWatermarkLightness{0.95}

\hypersetup{
	bookmarks=true,         
	unicode=false,          
	pdftoolbar=true,        
	pdfmenubar=true,        
	pdffitwindow=false,     
	pdfstartview={FitH},    
	pdftitle={My title},    
	pdfauthor={Author},     
	pdfsubject={Subject},   
	pdfcreator={Creator},   
	pdfproducer={Producer}, 
	pdfkeywords={keywords}, 
	pdfnewwindow=true,      
	colorlinks=true,       
	linkcolor=blue,          
	citecolor=blue,        
	filecolor=magenta,      
	urlcolor=cyan           
}

\usepackage[labelfont=bf,labelsep=period,justification=raggedright]{caption}

\makeatletter
\renewcommand{\@biblabel}[1]{\quad#1.}
\makeatother

\begin{document}

\title{We both think you did wrong - How agreement shapes and is shaped by indirect reciprocity - DRAFT}

\author{
Marcus Krellner$^{1,\star}$ and The Anh Han$^{1}$}

\maketitle
	{\footnotesize
		\noindent
		$^{1}$  School Computing, Engineering and Digital Technologies, Teesside University\\ 
		$^\star$ Corresponding authors: krellner.marcus@gmx.de
	}




\section*{Abstract}
Humans judge each other's actions, which at least partly functions to detect and deter cheating and to enable helpfulness in an indirect reciprocity fashion. However, most forms of judging do not only concern the action itself, but also the moral status of the receiving individual (to deter cheating it must be morally acceptable to withhold help from cheaters). 

This is a problem, when not everybody agrees who is good and who is bad. Although it has been widely acknowledged that disagreement may exist and that it can be detrimental for indirect reciprocity, the details of this crucial feature of moral judgments have never been studied in depth. 

We show, that even when everybody assesses individually (aka privately), some moral judgement systems (aka norms) can lead to high levels of agreement. We give a detailed account of the mechanisms which cause it and we show how to predict agreement analytically without requiring  agent-based simulations, and for any observation rate. Finally, we show that agreement may increase or decrease reputations and therefore how much helpfulness (aka cooperation) occurs.
 \vspace{0.2in}
 
 \noindent \textbf{Keywords:cooperation, evolutionary game theory, indirect reciprocity, donation game, private assessments, analytical predictions} 
 
\newpage

\section{Introduction}

From simplest replicators, evolution has worked its wonders and today we marvel at the success, diversity and complexity of life. Many wonders of life were enabled by cooperation. Cells work together to form multicellular organisms, former bacteria serve as the powerplants of cells in exchange for eternal protection from an outside world, and individual organism work together in communities of thousands and millions to become the most widespread forms of life there are, ants/wasps and, in similar but different way, humans.  

Indirect Reciprocity (IR) is one of the few other mechanisms to facilitate cooperation among self-interested agents \citep{Rand2013,Nowak2005,perc2017statistical}. You help me, without receiving any direct returns (that is, in that moment you behave altruistically). But you will be rewarded by others in the future. Because others have observed your behavior and will like you for it. I.e. they judge your action and form an opinion about you. And because they like you, they decide to help you. Indirect reciprocity allows cooperation (aka continued help amongst helpers) in large groups of unrelated individuals, even if they do not interact enough to establish direct reciprocity \citep{Schmid2022}. Hence, indirect reciprocity has been the focus of many influential studies of evolutionary game theory (\cite{Nowak1998,Ohtsuki2004a}, see also \cite{Sigmund}) and is even considered as an important foundation of human morality \citep{Nowak2005}.

Traditionally, studies simplified the mechanism of IR, assuming that judgments are unanimous (also called public assessments). More realistically, everybody would make their own judgements and therefore have their own opinions (aka private assessments). When the public assessment assumption is removed, dynamics and outcomes of IR models  change \citep{Brandt2004,Okada2020a}, often to the worse for the evolution of cooperation \citep{Uchida2013,Hilbe2018}. 
The main reason is the emergence and spreading of disagreements about someone's reputation, which had not been possible with public assessments. Disagreements can significantly disturb IR strategies that are stable under the public assessment regime, because they hinder a universal principle: if you withhold help towards somebody bad, this should not be judged as bad \footnote{at least if done by somebody good, which can be seen as in general judging it neutral or even good \citep{Okada2020a,Krellner2022}.} \citep{Panchanathan2003,Ohtsuki2006}. It is problematic to apply this rule, if potential helpers and observers may disagree. You withhold help to me, since you do not like me. But some observers like me, so they will disapprove of your action. So, if there was disagreement about my reputation, your reputation can get tainted.

In addition, a disagreement can cause further disagreements. If two observers disagree about me, they may also judge you differently for withholding help. That is, they now disagree about another person's reputation. Even a single disagreement can cascade through the population, even if no further errors occur, so that all opinions become bad or essentially random \citep{Hilbe2018}. 

Private assessment and disagreements are not only detrimental for IR strategies, but also lead to new challenges for IR research. Whereas for public assessment, there had long been exhaustive investigations of hundreds of strategies \citep{Ohtsuki2004a}[santos] using analytical models, something similar for private assessment was only achieved in recent years \citep{Okada2020a,Perret2021}. And these models still require restricting assumptions. Namely that, in an infinite population, only a finite number of players observe. That is, the probability to observe $\psi$ and hence also the fraction of opinions changed by a single interaction are required to be negligible ($\psi \rightarrow 0$).

Analytical models and most models of IR in general \citep{Okada2020b} consider opinions as binary. That is, an individual is judged as either good (1) or bad (0). To understand the state of opinions in a population, it is important to know the average reputation $r$: the probability of a random opinion being good, as well as the average probability of an agreement $a$: the probability that two randomly selected players have the same opinion about another random player.  

The models mentioned above only considered $r$ (hence we shall call them \textit{R-models} for short). They assume that, when the observation probability is very small ($\psi \to 0$), the probability of agreement $a$ only depends on $r$ and is given by $\widehat{a} = r^2 + (1-r)^2$ (i.e. probability that two random opinions are either both good or both bad) \citep{Okada2018}. 

\begin{figure}
\begin{center}
\includegraphics[width=1\linewidth]{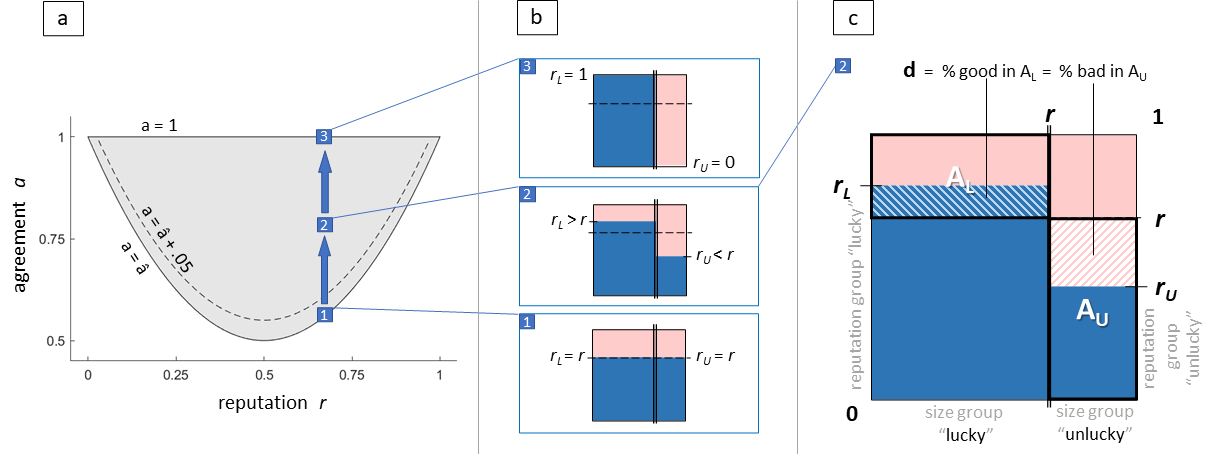}
\caption{\\
a) Two dimensions of global opinion state. \textmd{Graph shows agreement-reputation space, with the solid lines $a=1$ and $a=\widehat{a}$ indicating outer borders of possible agreement values in grey. (Subsequently, values above $a=\widehat{a}+0.05$ will be considered as significant additional agreement, since values very close to $\widehat{a}$ could conceivably be the result of noise or finite properties of the simulations; or be just too insignificant to make a difference). Case 1 has minimal agreement for the specific global average reputation $r$. Case 3 has maximal agreement for the same $r$ and case 2 an intermediate one for the same $r$ still.}\\
b) Model of simplified image matrices. \textmd{We simplify the image matrices for the three cases by removing all information about whose opinions it is and what specific player it is about. We only consider whether the opinion is about a player of group 1 or group 2 . The result can be imagined as two containers (left and right, separated by the double line) holding good opinions (dark blue areas). We may call the containers or groups "lucky" on the left and "unlucky" on the right. In case 1, both containers are filled to the same height, representing that the reputations of lucky and unlucky group, $r_L$ and  $r_U$, are equal to the global average. In case 2, some good opinions from the unlucky container were 'poured' into the lucky container, keeping $r$ unchanged since we do not add or remove any good opinions. In case 3, all remaining of these good opinions were poured as well, filling the lucky container exactly to the brim. It is apparent that in case 3, all players of the lucky group have only good opinions about them, therefore agreement about a player of this group is always $100\%$. The same applies to the unlucky group, which has only bad opinions about them.} \\
c) Size of the groups and precise state of agreement \textit{d}. \textmd{The complete transition from case 1 to case 3 can only be achieved under a specific condition. The size of the groups (the width of the containers) must be $r$ and $1-r$, respectively. With this, the areas $A_L$ and $A_U$ are of equal size. That means, for minimal agreement as in case 1, there are exactly as many good opinions about the unlucky group as there are bad opinions about the lucky one. Therefore, they can be entirely exchanged. The central feature of the analytical model introduced in this paper, is the parameter $d$, which describes the percentage of good opinions from area $A_U$ exchanged with bad opinions from area $A_L$. This parameter is independent of $r$ and can obtain any value in the interval $[0,1]$ (note that both areas disappear for the extreme cases $r=0$ and $r=1$, which is why $d$ is arbitrary in these cases). For details see equations \ref{rL} and \ref{a}.}}
\label{fig1}
\end{center}
\end{figure}

In theory, $a$ can be much higher. The population could for example consist of two groups, one with perfect reputation, the other with an abysmal one, see case 3 in Figure \ref{fig1}. The average reputation is still $r=0.6$. But all opinions about a player of the unlucky group are bad. Which means that the opinions of any two players about such a player will always be the same. This is similarly true for the lucky group in which all opinions are good. In case 3, average agreement is therefore 100\%. There are of course also intermediate states, such as case 2. The left side of the figure shows the entire agreement-reputation space and how agreement can range anywhere between $\widehat{a}$ and 1. Note that $0.5\leq\widehat{a}\leq 1$ (since $0\leq r \leq1$). 

There are obvious mechanisms that cause disagreement and push state of the population towards $\widehat{a}$, as discussed above. However, there could be other mechanisms that increase agreement. For example, if you donated your help, observers may not care about their current opinion about the recipient and like you either way (which is the case for image scoring \citep{Nowak1998a} or standing \citep{Sudgen1986,Leimar2001}). Hence, aside from a few who miss-perceive your action due to noise, all observers will have the same opinion about you. 

The existence of \textit{additional agreement}, i.e. states where agreement is significantly above $\widehat{a}$, was already implicitly shown by the work of \cite{Fujimoto2022}, who investigated the exact distributions of reputations for four strategies. They found that even infinite populations sometimes have two or even up to an infinite number of distinct reputation states, that a player could be in. It was not stated by the authors 
but any split of players into groups with different reputation states entails that the opinions of this population have additional agreement (see a proof sketch in Supplementary Information (\textbf{SI})). Nevertheless, their seminal work included an analytical model to predict distributions of strategies exactly, which will likely be foundational for future analytical models of IR. Yet, two problems remain. The number of strategies it can be applied to is small and the model is limited to the condition of full observation (i.e., $\psi=1$).

In conclusion, in the literature so far, there have been only two types of analytical models of  private assessment, and they are all limited to either $\psi=1$ or $\psi\to0$. Both cases are extreme ends of possible natural conditions. Full observation seems impossible due to physical constraints. Even if all interactions between people were public, there are still limitations to the attention observers can pay. Nominal observation rates, on the other hand, make IR less effective in the short-term, and may cause it to be replaced by direct reciprocity \citep{Schmid2021}. As \cite{Fujimoto2022} state themselves, a model for intermediate observation rates is necessary, especially to study IR under private assessment in an exhaustive fashion, like the defining work of \cite{Ohtsuki2004a}.

Overall, our investigation fills two major gaps in the literature. We provide the first the exhaustive investigation of agreement in reputation dynamics under private assessment. We test whether additional agreement is common for high or intermediate observation rates, but also if the assumption of the R-models always holds for $\psi\to 0$. We test this with simulations of a vast number of strategies and conditions. Second, we define the first \textit{A-R-model}, that moves beyond the one-dimensional approach of the R-models, to capture the two-dimensional space of agreement and reputation. It is capable of predicting long-term behavior of said simulations. Both contributions advance the understanding of indirect reciprocity under private assessment and lay important groundwork for exhaustive evolutionary investigations.

\section{Methods}

\subsection{Simulation}

We examine if the average agreement $a$ is greater than assumed in the R-model, $\widehat{a} = r^2 + (1-r)^2$. For that, we run simulations with different strategies and conditions. To limit complexity, following the example of \cite{Fujimoto2022}, we consider homogeneous populations, i.e. within a simulation all agents follow the same strategy. 

We have chosen a framework to model many different strategies. The first characteristic of a strategy is its assessment rules, given by vector $\alpha$. It defines how an observer judges the action, i.e. cooperation $C$ or defection $D$, depending on its opinion about the recipient, i.e. good $G$ or bad $B$. An assessment can be one of three forms: the action is approved of (+1), disapproved of (-1) or ignored (0). If the player approves of an action, their opinion of the donor will be good afterwards (which is the same as to say that it has increased within the limit of the two opinion values, 0 and 1, aka bad and good). If the action is disapproved of, the opinion will be bad. And if the action is ignored, the opinion does not change. 

The second characteristic of the strategy is given by its action rules $\beta$. They define what players will do, when they meet a recipient they like or dislike. These actions are cooperate (1) or defect (0). In the following, if we refer to $\alpha$ and $\beta$ as a whole, in the order given below. But we will sometimes refer to specific values or pairs of values, such $\alpha_{CG}$: cooperation towards good. With this framework we can model 324 different strategies. For example  the commonly studied strategy “staying” \citep{Okada2017,Sasaki2017} is described as follows:
\begin{equation}
\begin{matrix}
    staying: & 
    \begin{matrix}
         \alpha=\{\alpha_{CG},\alpha_{DG},\alpha_{CB},\alpha_{DB}\} = (1,-1,0,0) \\
        \beta=\{\beta_{G},\beta_{B}\} = (1,0) 
        \end{matrix}
        
\end{matrix}
\label{stay}
\end{equation} 
Of all the possible strategies of our framework, we study 171 strategies with unique behavior. Each of the possible strategies has a mirror image, which behaves equivalently in regard to cooperation. If we were to label opinion states as blue and red (instead of good and bad), an mirror image would cooperate in the exact same way to the same partners, but would think of them as blue instead of red or vice versa. We obtain such a mirror image of a strategy by exchanging the symbols +1 and -1 in the assessment rules and flipping them for good and bad recipients ($\alpha_m=(x_1,x_2,x_3,x_4)$ to $\alpha=(-x_3,-x_4,-x_1,-x_2)$) as well as flipping values of the action rules ($\beta=(xy)$ to $\beta_m=(yx)$). Note, that some strategies are their own mirror strategy. This focus on unique strategies conveniently leaves us with only three possible action rules: (1,1): unconditional cooperators or AllC, (1,0): conditional cooperators or Coco\footnote{sometimes, somewhat unfortunately, called discriminators} and (0,0): unconditional defectors or AllD (see SI for examples of mirror images). 

For our simulations, we consider a well-mixed population of size $N=100$. Every agent can have either a good (1) or a bad (0) opinion about each other agent. Hence, the state of the population can be described by the $N \times N$ image matrix $M(t)$ of the population at time $t$ \citep{Uchida}. Initially all entries are filled by a fair coin toss. Other studies have reported, that other initial conditions almost never change the outcomes \citep{Hilbe2018}. 

Time is discrete. In total, each time step consists of three parts. First, a donor $do$ and a recipient $re$ are drawn at random from the population. The donor then decides whether to cooperate. Note, these two steps are usually referred to as the donation game. To apply IR, donors base their decision on their action rule and their current opinion about the recipient. In the third part, opinions about the donor are updated due to observations.

These observations and updating can be broken down further. First, for each player (except the donor and recipient) it is decided if they will observe the interaction. For each player, who observes, it is determined if it observes accurately or if the observation is altered by a perception error, i.e. with a probability $\epsilon$ they will perceive the opposite of what the donor is actually doing. Next, individually perceived action and individual opinion of the observer about the recipient are combined for the private assessment of the donor. The opinion is updated if assessment is 1 or -1, but left as is if assessment is 0. Finally, each observer (whether they made an assessment or not) may change their opinion to a random value if they commit a cognitive error with probability $\mu$.

We study the behavior of the simulation in the long term and extract precise values for average reputation $r$ and average agreement $a$. In order to ensure accurate and reliable results, the simulations are run until we are reach an objective level of confidence about the measured average values (for details see \textbf{SI}.)

Note, in our version of the donation game, the donor and recipient do not update their opinion. I.e. players ignore the information of interactions they are themselves involved. This means we study pure indirect reciprocity, without any direct reciprocity \citep{Schmid2021}. Because of that, the diagonal of the image matrix is never updated nor used. We therefore exclude it from all computations of averages. For a detailed motivation of this design see \textbf{SI}.

\subsection{Predictive A-R-Model}

How could we define a new model which can represent both $r$ and $a$, but is otherwise as simple as possible? It has to be able to represent a continuous transition of $a$, from the baseline of the former prediction $\widehat{a}$ to its maximum 1 (see transition from 1 to 3 in figure \ref{fig1}). The transition must be possible for any $0\leq r \leq 1 $ while keeping $r$ constant. As mentioned, we could first nominally divide the population into two groups, lucky and unlucky. To increase agreement a minimal step, we could take away a random good opinion about an unlucky player and exchange with a bad opinion about a lucky player. I.e. we decrease the average opinion of unlucky players and increase the average opinion about lucky players. We could continue to do so until lucky players have the average reputation $r_L=1$ and unlucky players have the average reputation $r_U=0$, hence both agreement about lucky players $a_L$ and agreement about unlucky players $a_U$ is 1. 

We can do this only if the number of good opinions about unlucky players $n_{GU}$ is equal to the number of bad opinions about the lucky players $n_{BL}$. This is the case, if the size of the lucky group (or the percentage of lucky players) is equal to $r$, hence the size of the unlucky group is $1-r$. With this assumption, we may define a single value $d$, which is both the percentage of good opinions about unlucky players stolen and the percentage of bad opinions about lucky players replaced (see figure \ref{fig1}c). 
\begin{equation} 
\label{rL}
    r_L=r+d(1-r) \quad \& \quad   r_U=r-dr,\quad 0\leq d\leq1.
\end{equation}
With this model we define the two dimensional space with just two variables. We assume a specific distribution of reputations, namely that there are exactly two groups (in general there could be up to $N$ groups in a population of $N$ players). The goal of our model is not to represent the exact distribution. Just to represent $a$ as well as  $r$, for which the two-group scenario provides a minimal model. It allows us to to model all agreement states in all reputation states with a single parameter $d$. 

Both $r$ and $d$ can range between 0 and 1 (whereas the minimum of $a$ would depend on current $r$). Note that $d$ is undefined for $r=0$ and $r=1$ (see equation \ref{d}), but it does not need to be. In the case of $r=0$, $d$ vanishes from $r_U$ and the size of the lucky group is zero, hence $n_{GU}$ is zero for any $d$. The opposite is true for $r=1$ respectively. Besides these extreme cases, $d$ is given by
\begin{equation} 
\label{d}
d=\frac{2^{1/2}(r(1 - r)(- 2r^2 + 2r + a - 1))^{1/2}}{2r(1 - r)} = \left(1- \frac{ 1- a}{2r(1-r)}\right)^{1/2}, \ 0<r<1.
\end{equation}
This close form of $d$ is derived by solving the equation (\ref{a}) below for $d$, by replacing $r_L$ and $r_U$ with the equations in  (\ref{rL}). The global average agreement $a$ is given by the sum of agreements about members in each group discounted by the group size, i.e. $a=ra_L+(1-r)a_U$. The agreement for a group is derived by its average reputation, that is $a_L=r_L^2+(1-r_L)^2$ and $a_U=r_U^2+(1-r_U)^2$. Thus, 
\begin{equation} 
\label{a}
a=r(r_L^2+(1-r_L)^2)+(1-r)(r_U^2+(1-r_U)^2).
\end{equation}
We will now use $r$, $r_L$ and $r_U$ to predict the probabilities of events of our simulation, assuming an infinite population. To model a time step of the simulation, we need to distinguish between onetime events and repeated events. There are three events that happen only once per time step: picking a donor, picking a recipient and the decision of the donor. For these we need to know three kinds of probabilities for later computation. First, a pair of probabilities $q$ about the donor's group affiliation: $q_L=r$ (lucky group) and $q_U=1-r$ (unlucky group). Second, four probabilities $p$ representing the combinations of donor's choice and recipient's group. For example, $p_{CL}$, the probability that the donor cooperates and the recipient is in the lucky group (where $\beta$ is the set of the donor's action rules and $1-r_L$ is the chance the donor has a bad opinion of the recipient from the lucky group), is given as follows 
\begin{equation} 
p_{CL}=r\Big(r_L \beta_G+(1-r_L) \beta_B\Big)
\end{equation}
or in general
\begin{equation} 
\label{P}
\begin{matrix}
p_{jk}=q_k\Big(r_k c_{jG}+(1-r_k) c_{jB}\Big), \\
\text{where } j\in\{C,D\}, \ k\in\{L,U\}  \text{ and } 
c_{Cx}=\beta_x, c_{Dx}=1-\beta_x.
\end{matrix}
\end{equation}
Third, a 2-by-4 matrix $G$ showing how likely the opinion of the donor is good after assessment. It depends on the previous reputation of the donor's group $r_i$ and the four assessment rules $\alpha$ (see equation \ref{stay}). For example, since staying judges cooperation with bad recipient as neutral, $\alpha_{CB}=0$, the probability to be considered good is equal to the previous reputation, which is $r_L$ if the donor belongs to the lucky group
\begin{equation}
G_{L,CB}=r_L,
\end{equation}
or in general
\begin{equation}
\begin{matrix}
G_{i,mn}=
\begin{matrix}
    1, & \text{if} &   \alpha_{mn}=1 \\
    r_i, &  \text{if} &  \alpha_{mn}=0\\
    0, &  \text{if} &  \alpha_{mn}=-1 
\end{matrix},
\\
\\
\text{where } i\in\{L,U\}, m\in\{C,D\}, n\in\{G,B\}.
 \end{matrix}
\end{equation}
These are the probabilities that depend onetime events. There are also probabilities of repeated events, namely observations, since there can be many observers. Each observation includes the action the observer perceives and what opinion it had about the recipient. Hence there can be four kinds of observations, and we compute their probability $O$ for each of the four combinations of onetime events $p$ described in equation \ref{P}. For example, in the event that the donor cooperated with a lucky recipient, we can compute the probability $O_{CL,CG}$, that the observer has a good opinion about this lucky recipient and that the observer perceives the cooperation without a perception error $\epsilon$.
\begin{equation} 
O_{CL,CG}=r_L(1-\epsilon),
\end{equation}
or in general
\begin{equation} 
\begin{matrix}
O_{jk,mn}=r_i^*e, \\
\\
\begin{matrix}
    r_x^*=r_x & \text{if} & n=G \\
    r_x^*=1-r_x & \text{if} & n=B \\
    e=\epsilon & \text{if} & m\neq j \\
    e=1-\epsilon & \text{if} & m= j \\
\end{matrix}
\\
\\
\text{where } j,m\in\{C,D\}, k\in\{L,U\}, n\in\{G,B\}. 
\end{matrix}
\end{equation}
We then take into account observation rates $\psi$ and cognitive errors $\mu$, which are also repeated events. Each observation might not take place, leaving opinions unchanged. When it takes place, the resulting opinion may be altered by a cognitive error and its value would be reversed. We therefore adjust $G$ to $G^*$.
\begin{equation}
    G^*_{i,mn}=(1-\psi)r_i + \psi(G_{i,mn}(1-\mu)+(1-G_{i,mn})\mu).
\end{equation}
With these probabilities we can now compute the probability of the eight possible interactions, that is,  the combination of all three onetime events, as follows 
\begin{equation} 
\Pi_{ijk}=q_{i}p_{jk},
\end{equation}
and the expected reputation of the donor for each of these combinations
\begin{equation} 
\Gamma_{ijk}=\sum_{m,n}G^*_{i,mn}O_{jk,mn}.
\end{equation}
Finally, we can now compute the expected change of reputation 
\begin{equation} 
\Delta_r=\Big(\sum_{i,j,k}\Gamma_{ijk} \Pi_{ijk}\Big) -r
\end{equation}
and the expected change of agreement
\begin{equation} 
\Delta_a=\Big(\sum_{i,j,k}((\Gamma_{ijk})^2 + (1-\Gamma_{ijk})^2) \Pi_{ijk}\Big) -a.
\end{equation}
Note, as it was when computing agreement in the simulation, the instances of interaction have to be computed separately. Each of the eight possible interactions predicts a specific agreement about a donor's image. We compute each instance and only then compute the expected value of $\Delta_a$. Note also, that we do not assume that the donor can be attributed to the lucky or unlucky group after assessment. These groups are simplifications of the real image matrix that we use to describe its state with only two parameters. Information about the exact expected reputations of the donor and hence the agreement about it for the eight possible interactions is lost. Our model does not give exact predictions.

To validate our analytical model, we create a numerical algorithm to find the stable equilibria points in the $r \times a$ space (see \textbf{SI} for a detailed description). We use it to find the equilibria for all cases which we simulated and compare the results. For additional comparison we also use an analytical approach that models agreement always as $\widehat{a}$, aka a R-model \citep{Perret2021}, and compare the fit of both predictive models.   

\section{Results}

\begin{figure*}
\centering
\begin{subfigure}[t]{.48\textwidth}
  \centering
  \includegraphics[width=1\linewidth]{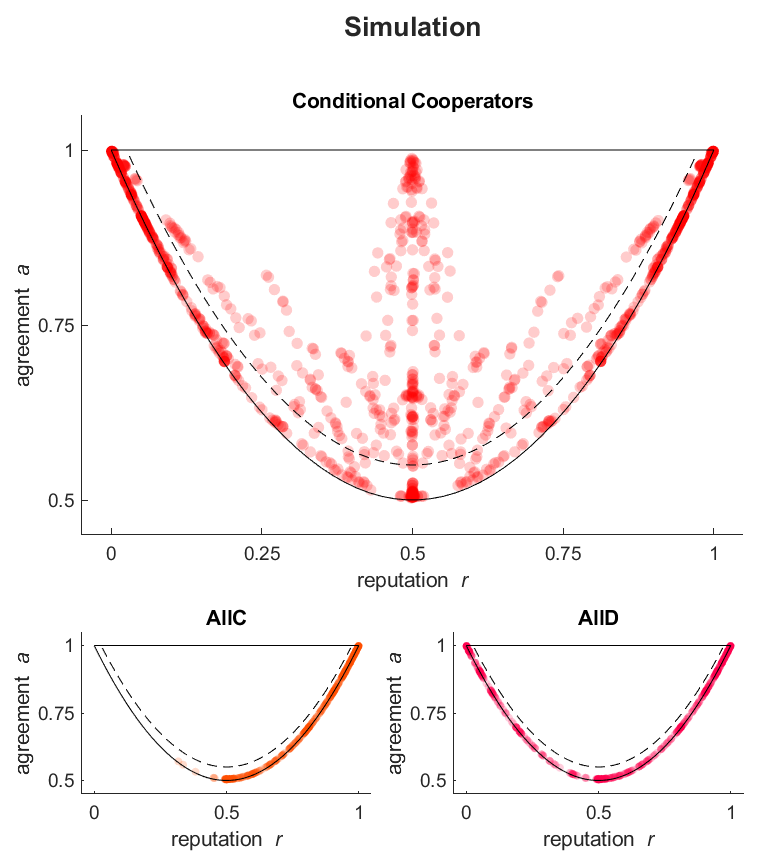}
  \caption{Simulations. \textmd{Agreement and reputation averaged across runs for homogeneous populations with 100 players. Not shown are cases where range of averaged reputations between runs exceeded 0.1  (2.9\% of cases for Cocos). For Cocos, $25.2\%$ have significant additional agreement (above the broken line).}
  }
  \label{fig:sub1}
\end{subfigure}\hfill
\begin{subfigure}[t]{.48\textwidth}
  \centering
  \includegraphics[width=1\linewidth]{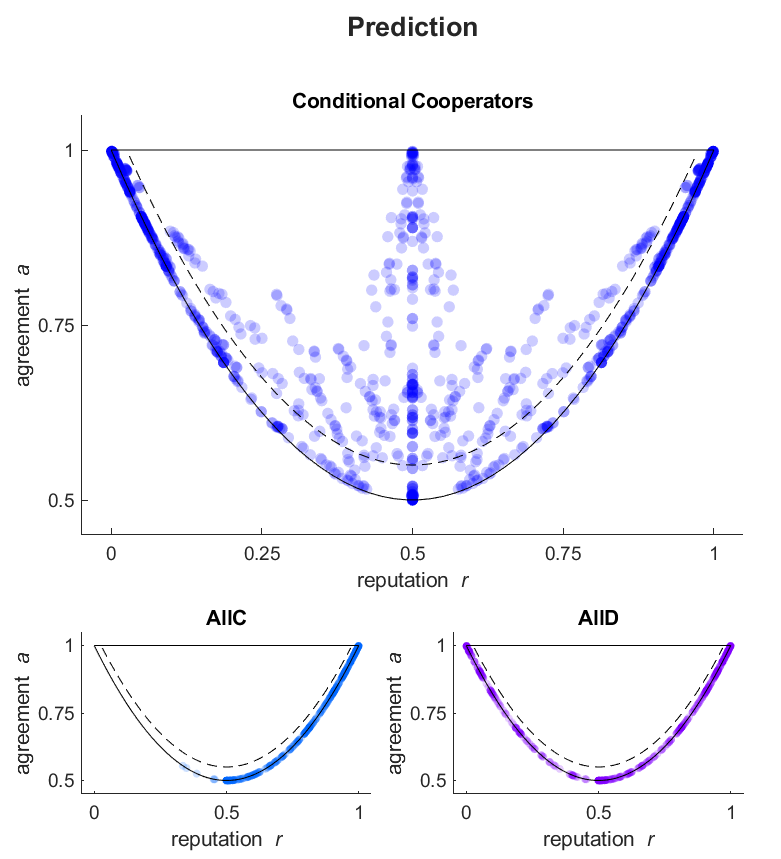}
  \caption{Predictions. \textmd{Unique stable states of the model for infinitely large homogeneous populations. Not shown are cases where multiple equilibria were found which ranged more than 0.01 in total distance in A-R space (for Cocos, 27 cases spread over only 3 strategies. See \textbf{SI} for some examples.)}
  }
  \label{fig:sub2}
\end{subfigure}%
\caption{Comparing results of simulations and analytical predictions. \textmd{Values for 171 unique strategies in 15 conditions (namely, $\epsilon \in \{0,0.001,0.01,0.05,0.1,0.2,0.4\}$ and $\mu \in \{0.001,0.01,0.05\}$). Reputation $r$ (x-axis) can range between $0$ and $1$. Agreement $a$ can range between $a=\widehat{a}$ and $a=1$ (solid black lines). Values above $a_{th}=\widehat{a}+0.05$ (broken black line) are considered as significant additional agreement.}}
\label{fig2}
\end{figure*}

\begin{figure*}
\centering
  \includegraphics[width=1\linewidth]{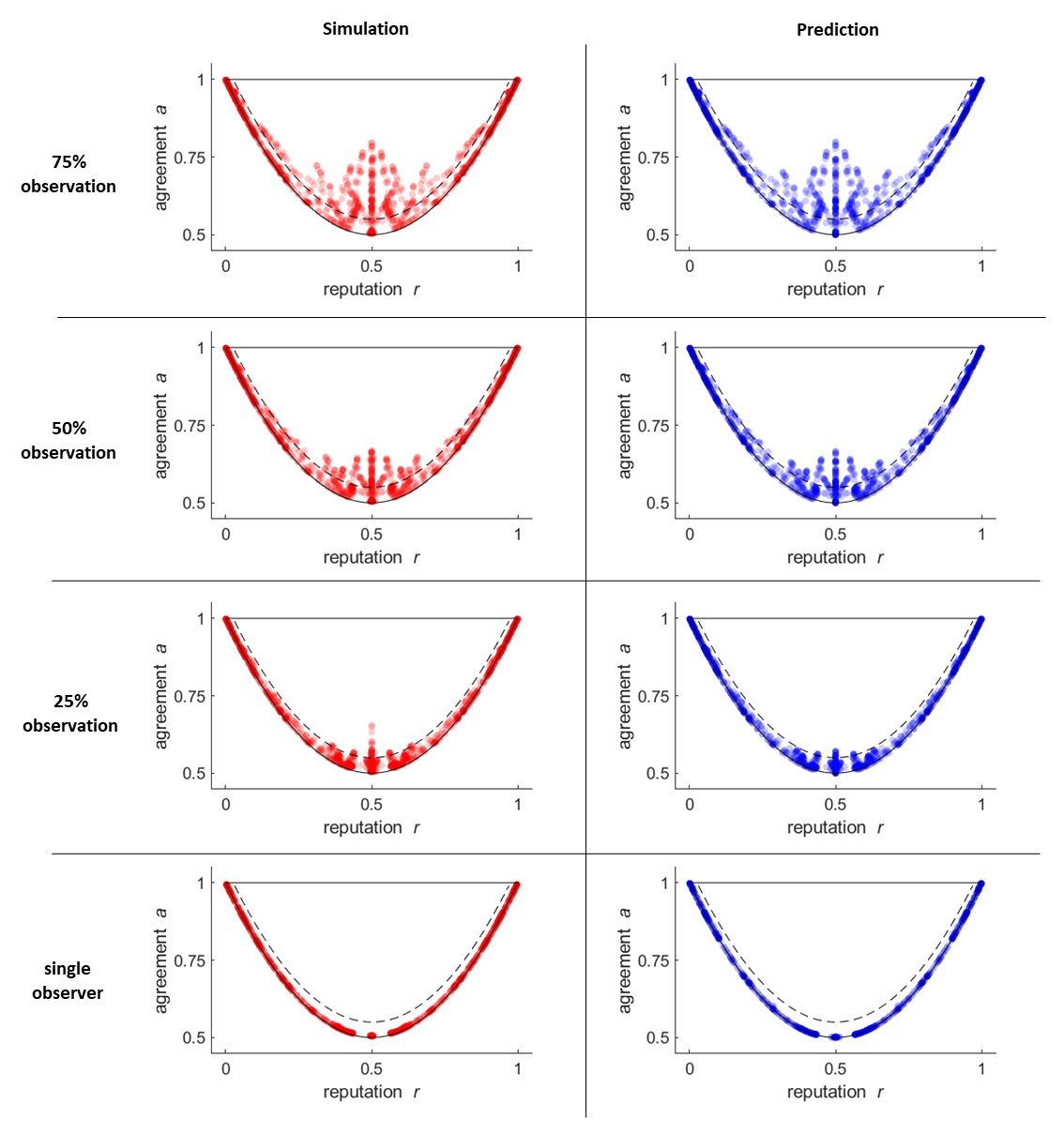}
\caption{Results for Cocos with other observation rates $\psi$. \textmd{On the left are the simulation results, and on the right, the predictions of the A-R-model. Values for all 81 unique conditional cooperators, with conditions and excluding criteria the same as those given in figure \ref{fig2}}.}
\label{fig3}
\end{figure*}

As described above, we studied 171 strategies with unique behavior. Each homogeneous population of 100 players is first tested for the observation rate $\psi=1$ in 15 conditions. The conditions span an exhaustive range of perception errors $\epsilon \in \{0,0.001,0.01,0.05,0.1,0.2,0.4\}$ and a reasonable range of cognitive error $\mu \in \{0.001,0.01,0.05\}$. Note, this full observation is the natural opposite to limiting assumption of the R-model where $\psi\to0$.

Figure \ref{fig:sub1} shows the results of our simulations as reddish dots. They are depicted separately for the three kinds as action rules: Coco, AllC and AllD. For AllC and AllD, agreement is always virtually minimal. But for Coco strategies, many points lay far above. Sometimes agreement is almost maximal ($a\to1$), while $\widehat{a}$ predicted it to be the smallest possible value ($a=0.5$ at $r=0.5$). For Cocos, 25.2\% of the measured agreements were at least 0.05 higher than expected (indicated by the broken black line). 

Note that we removed cases in which results differed significantly across runs. Averaging them could give a false impression of average agreement. For example, two runs with $r_1=0.3$ and $r_2=0.7$ and each minimal $a_i\to\widehat{a}=0.58$ would average to $r=0.5$ and $a=0.58$. For this averaged reputation, the minimal agreement would be $\widehat{a}=0.5$, so the average measurement would be $0.08$ higher than that. In other words, although the actual agreement in each run was minimal, the averaged agreement would indicate additional agreement above the threshold. We exclude cases in which the range of $r$ in a single condition is above $0.1$ ($2.8\%$ of cases in condition $\psi=1$, see \textbf{SI} for examples of excluded cases and alternative figure with cases included). It is clear that our findings are not caused by such artifacts and that for $\psi=1$ results differed substantially from the prediction $\widehat{a}$ with more agreement than expected for many conditional cooperators.

Next, we show that these results can be replicated for $\psi=1$ by our A-R-model, see  blue dots in figure \ref{fig:sub2}. Excluded are again cases for with multiple equilibria (given the much higher precision of the analytical approach we excluded all cases in which the total distance in the agreement-reputation space, similar to equation \ref{delta}, exceeded 0.01, see \textbf{SI} for examples of excluded cases). The general patterns of the simulation could be replicated. AllCs and AllDs have virtually no additional agreement whereas Cocos do. The qualitative fit between simulation and model is striking. 

Before we quantitatively describe the fit, we expand our investigation to other observation rates. We show only results for Cocos, since these are the most interesting cases, in figure \ref{fig3}. We studied three intermediate observation rates $\psi\in\{.75, .5, .25\}$ as well as a special case, where every round would have exactly one observer. A single observer is the smallest amount of meaningful observation possible. No observation would simply not change the image matrix and have no impact on reputation or agreement. It is the closest one can get to the limiting assumption of the R-Model $\psi\to0$.\footnote{Even taking a very small probabilistic observation rate such as $\psi=0.0001$ would have actually increased the number of observers. All rounds in which no observation takes place are simply ignored. The rest may have one or more observers. In addition, simulating many rounds just to be ignored would increase the computational demand unnecessarily for no benefit.} 

Both simulation and prediction show a steady decline in agreement, until $a$ is virtually minimal when there is only a single observer. This is a qualitative fit of what the R-model assumes. Adjusting the observation rate in the A-R-model replicates it. For all observation rates, qualitative fit is still high. And, similarly to full observation, intermediate observation rates still show significant additional agreement. 

We quantified the fit of simulation and predictions in two ways. Our A-R-model makes predictions about agreement $a_p$ as well as reputation $r_p$. We therefore compute the absolute distance to simulated agreement $a_s$ and reputation $r_s$ by 
\begin{equation}
    \Delta=\sqrt{(a_p-a_s)^2 + (r_p-r_s)^2}.
    \label{delta}
\end{equation} 
We excluded values not shown in figures \ref{fig2} and \ref{fig3} (if either simulation or prediction had to be excluded, we excluded the entire pair). Results for all observation rates are shown in the left of table \ref{tab1}. For both maximal and intermediate observation range, the fits are excellent. The median of absolute distance between the simulations and the model is 0.003. For 92.7-95.7\% of predictions, the deviation is smaller than 0.01. For the minimal observation rate, fits are slightly worse. The median 0.005 and only 72.3\% of predictions are less than 0.01 of target. 

We also compared the predictions of our A-R-model with predictions of the R-model. Since the R-model can only predict reputation $r_o$, we compared the two models in that regard. The difference in deviation is shown in the right of table \ref{tab1}. Keep in mind that the R-model was designed only for $\psi \to 0$, which is most closely met in the single observer condition. The new A-R-model is not often better as the R-model in this case. And at least for one case, it is much worse, as indicated by the range. However, the A-R-model is superior for all high and intermediate observation rates $\psi$. Its predictions are better in $4.6\%$ to $11.5\%$ of all cases (AllD, AllC and Cocos), and the accuracy is increased by as much as $0.5$, which is half of the maximum range of $r$. Higher accuracy is especially prevalent for Cocos, to which all evolutionary successful strategies belong, such as image scoring \citep{Nowak1998} and the leading-eight \citep{Ohtsuki2006}. Here, for $\psi=1$, the A-R-model is better in $24\%$ of cases. Considering an independently changing average agreement seems to have increased the prediction of average reputation substantially. 

\begin{table}
\centering
\caption{Deviations between Model Predictions and Simulation}
\label{tab1}
\begin{tblr}{
  width = \linewidth,
  colspec = {Q[200]Q[87]Q[81]Q[81]Q[163]Q[165]Q[160]},
  row{odd} = {c},
  row{2} = {c},
  row{4} = {c},
  row{6} = {c},
  row{8} = {c},
  cell{1}{2} = {c=3}{0.249\linewidth},
  cell{1}{5} = {c=3}{0.474\linewidth},
  cell{2}{2} = {c=3}{0.249\linewidth},
  cell{2}{5} = {c=3}{0.474\linewidth},
  hline{1,9} = {-}{0.08em},
  hline{4} = {-}{},
}
                & A-R-model & & & Difference A-R-model minus R-model & & \\
                & absolute~deviation & & & deviation in reputation & & \\
                & median                 &  $<.01$  &  $<.05$  & {$< -.01$ \\(A-R better)}                   & { $> .01$ \\(A-R worse)} & range        \\
$\psi = 1$      & 0.003                  & 92.7\% & 99\%   & 11.5\%                                       & 0\%                      & -0.498:0.006 \\
$\psi = 0.75$   & 0.003                  & 95.7\% & 99\%   & 9.7\%                                        & 0.1\%                    & -0.497:0.051 \\
$\psi = 0.5$    & 0.003                  & 95.4\% & 98.9\% & 7.8\%                                        & 0.2\%                    & -0.495:0.053 \\
$\psi = 0.25$   & 0.003                  & 94.7\% & 98.4\% & 4.6\%                                        & 0.1\%                    & -0.491:0.014 \\
single observer & 0.005                  & 72.3\% & 86\%   & 1.2\%                                        & 0.4\%                    & -0.034:0.334 
\end{tblr}
\end{table}

\section{Discussion}

We now summarize our findings and their immediate implications, before highlighting remaining issues and limitations. We further compare our A-R-model with the recent model of \cite{Fujimoto2022} to highlight the differences and common ground. We then discuss some important potential extensions of the A-R-model, which would enable it to implement recently proposed enhancements to IR strategies, such as generous assessment \citep{Schmid2021} and pleasing \citep{Krellner2021}. We will close with possible real world implications of the discovered additional agreement. 

In the first part of our report, we showed that additional agreement can emerge in indirect reciprocity under private assessment. For high or intermediate observation rates, opinions about a person were shared much more often than mere chance. Therefore, the assumption of the minimal agreement $\widehat{a}$, on which most previous models (R-models) relied, cannot be generalized to these circumstances. We could on the other hand confirm that the assumption is reasonable for very low observation rates, such as a single observer for each interaction. But we showed, that results about evolutionary success with solitary observers \citep{Okada2018,Okada2020a,Perret2021} cannot yet be generalized to other observation rates.

The second part of our report makes an important step towards that goal. Previous R-models can only model average reputation and therefore must rely on the assumption of minimal agreement. Our A-R-model is able to represent average levels of agreement and reputation independently. It predicts both with astounding accuracy and does so for any intermediate or high observation rates. It outperforms the predictions of R-models in these circumstances by a large margin.

Making precise measurements or predictions of reputation in particular is the basis for studies on the evolutionary stability of any reputation-based IR strategies \citep{Okada2020b}. An individual's reputation directly corresponds to how much cooperation they receive, hence it determines their payoffs and even the payoffs of others (from which the may receive cooperation, hence cause costs to them). 

Parallel to this work, other researchers have discovered another way to predict opinion dynamics more accurately than by average reputation alone. \cite{Fujimoto2022} were able to predict precise distributions of reputations, e.g. 80\% of players who would have a reputation of about 0.8, 15\% of 0.2, and 5\% of 0.25. This was another hugely important step for analytical models of IR under private assessment and is, of course, closely related to the current paper. 

Our model assumes a simplified distribution of reputations. We treat the populations as if there were at most two groups, and as if their size was given by the average reputation. \cite{Fujimoto2022} show that this is often not the case, that there can be more groups or other compositions. Their analytical approach is exact and uses no simplification. However, as the fit of our model shows, this level of detail may not be necessary. 

And the level of detail in their model seems to come at a cost as their analysis is limited to only four strategies. It seems reasonable however that their framework could model all strategies, that care about action as well as reputation of the recipient, but apply only binary assessment rules (i.e. assess each action as either good or bad, but not as neutral). This would cover 64 strategies (mirror images included) compared to the 324 of our study.\footnote{Indeed, in a preprint, \cite{Fujimoto2023} modelled 16 strategies, focusing on a single action rule, instead of four. If they had looked at all actions rules and not exclude mirror images, they could have modelled 64 strategies. However, since they were only concerned with levels of cooperation, the limitation to Cocos ($\beta=(1,0)$) was perfectly justified. AllD and AllC with different assessment rules may held different reputations, yet the all cooperate the same.} 

Their model also covers only a single observation rate $\psi=1$. The authors themselves state that being able to model arbitrary observation rates would be a very important extension of their approach. Our A-R-model is already able to do that. And, it seems at least possible, that their approach is not capable of dealing with intermediate observation rates, i.e. $\psi < 1$. Because their model currently relies on the fact that the new reputation of the donor is entirely independent of their current reputation. If some players do not observe, their new opinions depend (entirely) on their current opinions, since their opinions are just kept as they are. Their model could no longer be based on one dimension (the current reputation of recipients) but would have to incorporate entirely new dimension (the current reputation of the donor). This increase in complexity might be a serious problem.

These problems for arbitrary observation rates also concern the modelling of more complex, so-called 3rd-order strategies \citep{Santos2021}, which also use the current reputation of the donor in their assessment. In contrast, in our A-R-model, the current reputation of the donor is already incorporated. It will be straightforward to extend our model to consider all 2048 3rd-order strategies (mirror images included) \citep{Ohtsuki2004}.


We envisage the following additional extensions of our model. First, we can include strategies that do not use deterministic rules but probabilistic ones \citep{Schmid2021,Schmid2021a}. Instead of assessing defection always as bad, observers may only do so 80\% of the time (in other words they are sometimes generous in the judgment). The same could be applied to their action rules. They may want to cooperate even with bad individuals about 10\% of the time. Second, we can include pleasing \citep{Krellner2020,Krellner2021}. Instead of granting or refusing help only based on the donor's own opinion (which can be easily disturbed by perception error), the donor pools some of the opinions of others and act like the majority would decide. Some of these extensions are as easy as replacing -1, 0 or 1 in the action and assessment rules of this paper with -0.8 and 0.1. 


\begin{figure*}
\begin{center}
\includegraphics[width=1\textwidth]{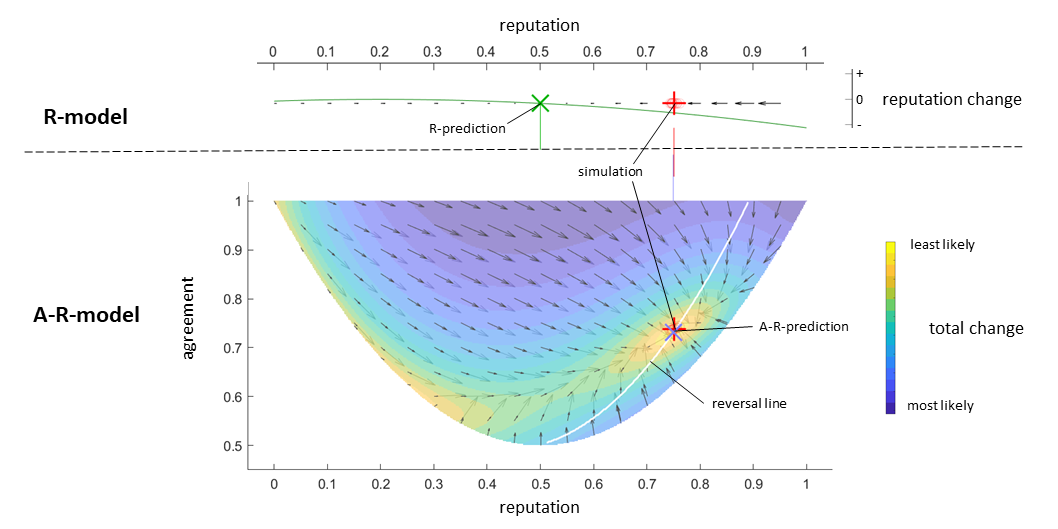}
\caption{Detailed comparison between models for staying norm \textmd{(see equation \ref{stay}). 
Parameters are $\psi=1$, $\epsilon_p=0.1$ and $\mu=0.01$. Bottom graph shows the two-dimensional space of the A-R-model. Dark blue areas indicate likely changes in either agreement or reputation or both, whereas light yellow areas indicate little to no change. Arrows indicate the direction of the change. The precise predictions of the model is indicated by a blue x. Simulations are shown in red, the averages of single runs are shown as red circles and their total average as a red +. Top graph shows one-dimensional R-model in comparison. Change in reputation and its direction is depicted on the y-axis in addition to the arrows. The stable point (green x) is where the x-axis is crossed with a downward slope. The white line in the A-R-model indicates a shift from reputation decrease (below) to reputation increase (above). }}
\label{fig4}
\end{center}
\end{figure*}

This paper is the first to study agreement in such detail. Agreement is the key feature that distinguishes previous  research on IR under public assessment from the more recent research on private assessment. Public assessment fixes agreement at its maximum (all agree all the time), but private assessment does not fix it at the opposite state i.e. its minimum $\widehat{a}$. Public and private are not opposite sides of the same coin. Rather private assessment allows agreement to vary, opening a new dimension of complexity to the dynamics of reputation-based IR. 

Our A-R-model allows one to study agreement and reputation in even more detail than was reported within the results of this paper. We focused on the stable states of the population. But one could study the direction and likelihood of change in each state, as seen in figure \ref{fig4} for the example of the staying strategy (see also \textbf{SI} for other important strategies, such as image scoring \citep{Nowak1998} and the leading-eight \citep{Ohtsuki2006}). One important insight this grants is about the stability of other regions. For staying, there exist somewhat stable areas with reputations lower than $r=0.3$ and minimal agreement. The existence of additional relatively stable regions has large implications for the short-term behavior of the population.

Seeing the direction of change in the entire space also allows us to make educated guesses about another dynamic. The white line in the A-R-model of figure \ref{fig4} indicates a shift from reputation decrease (below) to reputation increase (above). This highlights how the trend in reputation depends on the state of agreement. It seems reasonable, that increasing agreement could increase reputation as well. Increase in agreement could be done by a form of opinion synchronization, for example by some players gossiping about their observation or opinions to make other opinions fall into line with their own. \cite{Hilbe2018} suggested that any IR strategy would profit of, or indeed rely on,  some sort of opinion synchronization to maintain stable cooperation under private assessment. With the A-R-model, we can visualize which strategies can actually profit from such mechanisms. Some are not affected, e.g. image scoring, some actually show the opposite pattern, such as "GKGB" \citep{Okada2020a}, which seems only stable under private assessment, but not under public one (see \textbf{SI} for examples). In the future, we can seek a way to incorporate a probability for opinion synchronization in the A-R-model, to predict new stable points under various synchronization regimes. 

Related to that is a last alteration of the model. In the current investigation, we considered private assessment in accordance with most literature \citep{Okada2020b}. Every individual or player in the population observes independently. But we can imagine a situation, where only a few players observe, who then share their assessment with many others. Consider for example the extreme case of a single observer that shares with the entire population. This would correspond to public assessment \citep{Ohtsuki2004a}, where no disagreement is possible ($a=1$). In a more realistic scenario, multiple observers are at least possible, and they may judge the interaction differently due to individual errors or different previously held opinions. The information of their judgments may also fail to reach some players. In such scenarios, disagreement can exist. However, all players who got the news from the same observer will most likely have the same opinion (even if information transfer is noisy). It is a form of built-in opinion synchronization. We can therefore expect agreement to be higher than in the scenario of the current paper. The amount of agreement reported here might only be the lower end for private assessment scenarios.  


Lastly, we look at the reasons additional agreement emerges. As discussed above, there can be no additional agreement if every player has the same reputation. Additional agreement requires at least two groups exist (but up to $N$-many), with different average reputations. Such groups can only emerge, if players with the same strategy face different fates. For example, a donor of strategy $\alpha=(1,-1,1,-1)$ and $\beta=(1,0)$ (image scoring) can get lucky by meeting a recipient they believe to be good. Hence the donor cooperates and earns a high reputation (everybody who observes, and does not make an error, believes the donor to be good). But another donor with the same strategy might get unlucky, because they believe the recipient to be bad. Hence they defect and most observers will now think this donor is bad. 

In general, groups can emerge if different onetime events cause different expected reputations. Such onetime event are which recipient is met or which decision is made. In the image scoring example, if errors happen with probability $0.1$, we expect the reputation of a donor to be either $0.1$ if they defected or $0.9$ if they cooperated. Different decisions as donor seem indeed to be the most important factor for additional agreement. That is why only conditional cooperators (Coco) show additional agreement, but unconditional cooperators (AllC) or defectors (AllD) show none. Because some Cocos cooperate and some defect, they can form groups with different reputations and therefore have additional agreement. 


Is additional agreement a good thing? Our results for strategies such as staying show that it can increase reputations. The reputations studied here correspond to how often a player finds another player of the same strategy worth of cooperation (and Cocos also act on it). More agreement can increase reputations, hence increase cooperation rates within the strategy, hence increase the stability of that strategy. This could keep defectors at bay and increase cooperation in general. Evolutionary investigations need to confirm these assumptions, but in that regard, agreement might be a good thing.

However, we also showed another consequence of agreement. It is connected with differences in reputations. Such differences inevitably lead to short-term inequality between players. Over the long run, each player of the same strategy will alternate between being lucky and being unlucky, so each will earn the same pay-offs. But some players may earn much less than others if only a few consecutive interactions are considered. This can be a problem, if the game (or life) would require a player to earn a minimal amount to sustain themselves. It may also be a problem if such inequality itself has bad consequences for individuals or society. 


Research on IR seems to continue to converge \citep{Okada2020b,Krellner2022}. However, there seem to be still a lot of potential to deepen our understanding of the processes, as demonstrated in \cite{Fujimoto2022} and the current paper. The understanding of agreement is central for any analytical model of IR under private assessment. It is central in understanding what benefits or problems that opinion synchronization or different observation rates may bring. And it is as central to all research on reputation-based IR under private assessment, since agreement can change independently of reputation and can significantly alter the latter as well. Our A-R-models provides the means to study the evolutionary dynamics of indirect reciprocity under private assessment for yet the largest strategy space and widest range of conditions.





\end{document}


\maketitle
\author{
M. Krellner$^{1}$ and T. A. Han$^{1}$
}

$^{1}$School of Computing, Engineering and Digital Technologies, Teesside University, UK\\

krellner.marcus@gmx.de

\section{What mirror images of strategies were excluded}

Each strategy of our framework has a mirror image, which behaves equivalently in regard to cooperation. If we were to label opinion states as blue and red (instead of good and bad), a mirror image would cooperate in the exact same way to the same partners, but would think of them as blue instead of red or vice versa. We obtain such a mirror image of a strategy in three steps (see table \ref{tabS1}). We exchange the symbols +1 and -1 in the assessment rules (a) and flip them for good and bad recipients (b), transforming $\alpha=(x_1,x_2,x_3,x_4)$ to $\alpha_m=(-x_3,-x_4,-x_1,-x_2)$. We also flip the values of the action rules (c), transforming $\beta=xy$ to $\beta_m=yx$. For the strategy staying, see mirror image in \ref{tabS1}.d). Sometimes the transition of assessment rules creates the same assessment rules again (e). If the action rule is either (0,0) or (1,1), such a strategies is its own mirror image (f). 
 Deciding which mirror images to keep is somewhat arbitrary. We decided to keep all $\beta=10$ strategies and excluded their mirror images, which eliminated the action rule $\beta=01$ entirely from the population of strategies. This action rules suggest cooperating with bad players and defecting towards good players, which seem intuitively wrong. We were left with only three possible action rules: (1,1): unconditional cooperators or AllC, (1,0): conditional cooperators or Coco and (0,0): unconditional defectors or AllD.
 
\begin{table}[h]
\begin{tabular}{l|ccccccc|lcccccc|}
\cline{2-15}
\multirow{3}{*}{}        & \multicolumn{6}{c}{\textbf{strategy}}                                                                                                                                                          & \textbf{} &  & \multicolumn{6}{c|}{\textbf{mirror image}}                                                                                                                                                     \\ \cline{2-15} 
                         & \multicolumn{2}{c|}{action rules}                               & \multicolumn{4}{c}{assessment rules}                                                                                         &           &  & \multicolumn{2}{c|}{action rules}                               & \multicolumn{4}{c|}{assessment rules}                                                                                        \\ \cline{2-15} 
                         & \multicolumn{1}{c|}{$\beta_G$} & \multicolumn{1}{c|}{$\beta_B$} & \multicolumn{1}{c|}{$\alpha_{CG}$} & \multicolumn{1}{c|}{$\alpha_{DG}$} & \multicolumn{1}{c|}{$\alpha_{CB}$} & $\alpha_{DB}$ &           &  & \multicolumn{1}{c|}{$\beta_G$} & \multicolumn{1}{c|}{$\beta_B$} & \multicolumn{1}{c|}{$\alpha_{CG}$} & \multicolumn{1}{c|}{$\alpha_{DG}$} & \multicolumn{1}{c|}{$\alpha_{CB}$} & $\alpha_{DB}$ \\ \hline
\multicolumn{1}{|l|}{a)} & \multicolumn{1}{c|}{}          & \multicolumn{1}{c|}{}          & \multicolumn{1}{c|}{1}             & \multicolumn{1}{c|}{1}             & \multicolumn{1}{c|}{1}             & 1             &           &  & \multicolumn{1}{c|}{}          & \multicolumn{1}{c|}{}          & \multicolumn{1}{c|}{-1}            & \multicolumn{1}{c|}{-1}            & \multicolumn{1}{c|}{-1}            & -1            \\ \hline
\multicolumn{1}{|l|}{b)} & \multicolumn{1}{c|}{}          & \multicolumn{1}{c|}{}          & \multicolumn{1}{c|}{1}             & \multicolumn{1}{c|}{0}             & \multicolumn{1}{c|}{0}             & 0             &           &  & \multicolumn{1}{c|}{}          & \multicolumn{1}{c|}{}          & \multicolumn{1}{c|}{0}             & \multicolumn{1}{c|}{0}             & \multicolumn{1}{c|}{-1}            & 0             \\ \hline
\multicolumn{1}{|l|}{c)} & \multicolumn{1}{c|}{1}         & \multicolumn{1}{c|}{0}         & \multicolumn{1}{c|}{}              & \multicolumn{1}{c|}{}              & \multicolumn{1}{c|}{}              &               &           &  & \multicolumn{1}{c|}{0}         & \multicolumn{1}{c|}{1}         & \multicolumn{1}{c|}{}              & \multicolumn{1}{c|}{}              & \multicolumn{1}{c|}{}              &               \\ \hline
\multicolumn{1}{|l|}{d)} & \multicolumn{1}{c|}{1}         & \multicolumn{1}{c|}{0}         & \multicolumn{1}{c|}{1}             & \multicolumn{1}{c|}{-1}            & \multicolumn{1}{c|}{0}             & 0             &           &  & \multicolumn{1}{c|}{0}         & \multicolumn{1}{c|}{1}         & \multicolumn{1}{c|}{0}             & \multicolumn{1}{c|}{0}             & \multicolumn{1}{c|}{-1}            & 1             \\ \hline
\multicolumn{1}{|l|}{e)} & \multicolumn{1}{c|}{1}         & \multicolumn{1}{c|}{0}         & \multicolumn{1}{c|}{1}             & \multicolumn{1}{c|}{0}             & \multicolumn{1}{c|}{-1}            & 0             &           &  & \multicolumn{1}{c|}{0}         & \multicolumn{1}{c|}{1}         & \multicolumn{1}{c|}{1}             & \multicolumn{1}{c|}{0}             & \multicolumn{1}{c|}{-1}            & 0             \\ \hline
\multicolumn{1}{|l|}{f)} & \multicolumn{1}{c|}{1}         & \multicolumn{1}{c|}{1}         & \multicolumn{1}{c|}{1}             & \multicolumn{1}{c|}{0}             & \multicolumn{1}{c|}{-1}            & 0             &           &  & \multicolumn{1}{c|}{1}         & \multicolumn{1}{c|}{1}         & \multicolumn{1}{c|}{1}             & \multicolumn{1}{c|}{0}             & \multicolumn{1}{c|}{-1}            & 0             \\ \hline
\end{tabular}
\caption{Examples for mirror images. \textmd{}}
\label{tabS1}
\end{table}



\section{Adaptive algorithm to run simulations until precise average values are reached}

We break our simulations into multiple parts and after each decide whether to continue or not. The smallest unit is a segment consisting of $10^3$ time steps. After a segment, we compute the reputations and agreements about all individual players at all time steps, and then average them over time and population. The average agreement $a_i(t)$ about a player $i$ at time $t$ is computed by a similar formula as before, $a_i(t) = r_i^2(t) + (1-r_i(t))^2$, i.e. it is given by their reputation at that time $r_i(t)$. They are computed individually, since  averaging reputation first would destroy the information about any agreement above $\widehat{a}$. 

After computing $r$ and $a$ of the segment, we average the values of the last half (rounded down to closest integer) of segments to estimate long term $r$ and $a$ (i.e we include middle one if uneven number and the fist estimation is just the first value). Only after at least 10 segments, we compute the standard deviations of the last 10 estimates. We stop the simulation if the standard deviations of both estimates are below 0.001. This constitutes a run and the last estimation will be saved as its result.

For the next run, we initiate the image matrix with new random opinions. Since we do not know, if the populations will show the same trends for each run, we again continue the simulations until we reach a confident estimation. After each run we compute a new estimation of $r$ and $a$ by averaging all runs (not just half). If, after at least 10 runs are done, the standard deviation of the last 10  estimations is below 0.001, we stop further runs and continue with the next condition. The results of each run of an condition and their averages are saved. For example, the simulations for condition $\psi=1$ ran on average for 19.9 segments each run (i.e. 19900 time steps) and were repeated for 13.4 runs until they reached a satisfying accuracy.

\section{Why Donor and Recipient do not update their opinion about the Donor}

Indirect Reciprocity (IR) is often referred to as a different mechanism from direct reciprocity (DR) \citep{Rand2013}. However, as was shown by \cite{Schmid2021}, both can exist in the same framework. In their model, DR is just a special case of IR, where the probability to use information outside ones' own interaction is zero. They assumed that observer and recipient always observe their own interactions. Hence, if they do not observe any other interactions, they judge others only by how they were treated themselves. The strategy, which evolves, closely resembles generous tit-for-tat \citep{NowakMartinA.Sigmund}.

Two problems arise if we let donors and recipients judge under private assessments. First, a donor judges itself systematically different from the average of how it is judged by others. Others may have a different opinion about the recipient than the donor (agreement is almost never perfect). Hence some will agree with the donor and judge it like the donor judges itself, but others will not. The donor is systematically biased about itself compared to the rest of the population. (For strategies such as "staying" where the donor plays the optimal action rule for its norms, the donor will tend to have a better opinion about itself than the rest of the population has.) 

Every player has this self-bias, since they only ever judge themselves while being a donor. Because of this bias, judgments by recipients are biased as well. A recipient judges actions according to its opinion about itself. Having an systematically different opinion about oneself causes systematically different opinions about donors while being recipient. 

For our investigation in particular, this would have been a problem, since the impact of this bias systematically differs between finite and infinite populations, which we directly compared. In infinite populations, self-bias is not a factor, since any finite number of opinions has no impact on the total average of opinions. However, it can have an impact in finite populations.

Furthermore, there is the problem that the impact is not constant for different observation rates $\psi$. In finite populations, where recipient and donor always judge, but only one other player does, the impact of the judgements of donor and recipient will be much higher than if 98 other players judge at the same time. This concerns the impact of self-bias, but also the impact of direct reciprocity \cite{Schmid2021}. We considered both to be undesirable for the current study, and therefore excluded donor and recipient from judging. Judgements as a donor are only meaningful if they are later used in judgments as a recipient anyway. Our approach excluded self-bias and DR, and allowed the dynamics of finite and infinite populations to resemble each other even for very small observation rates.

\section{Numerical algorithm to find equilibria of A-R and R-models}
\label{numAlg}

Given the definitions above, we can define a function $f$ which takes the inputs: current reputation $r$, current agreement $a$, as well as the description of the strategy $\alpha$ and $\beta$ and the two error parameters $\epsilon$ and $\mu$. And, which outputs are the expected change $\Delta_r$ and $\Delta_a$. So, we can determine whether reputation and agreement are expected to increase or decrease for a current state. We now look for states where both expected changes are zero, i.e. equilibrium points. These states are candidates for stable points, which predict the state of the system after a long period of time.

Finding equilibria for our models is difficult. The above-mentioned function cannot be solved symbolically, since the degree of polynomials in general is rather large. We created a two-level approach to find the equilibria numerically. On the lower level we use the function with the "solve" of Matlab2022b library \citep{Inc.2022}. The function has strengths and weaknesses. It uses a deterministic algorithm and is very fast and precise. However, it requires the input of a starting point and may fail to find the equilibrium from this point if it runs in local minima or, as is the case for our function, if it runs in undefined states (the function is not defined outside of $0 \leq r\leq 1$ and $r^2+(1-r)^2\leq a\leq 1$). 

We therefore add a second layer. We try multiple starting points and weed out failed attempts after the fact. We have chosen enough starting points (namely, 101) to find at least one equilibrium for each cases, and use the same starting points for all of them. This would sometimes yield multiple equilibria. Such results fall into two categories. Either the points lay closely together (total distance <0.01), which indicates a single true equilibrium that could not be computed with the same precision from all directions. In that case we look for the result with the smallest output of $f$ and discard the others. In the second category, there are indeed multiple numerically derived equilibria, some of which may not be stable. This is the case for just 3 out the 171 unique strategies. To simplify further analysis, we exclude these from the comparison with the simulations.

\bibliographystyle{apalike}


\begin{figure*}[t]
\begin{center}
\includegraphics[width=0.6\textwidth]{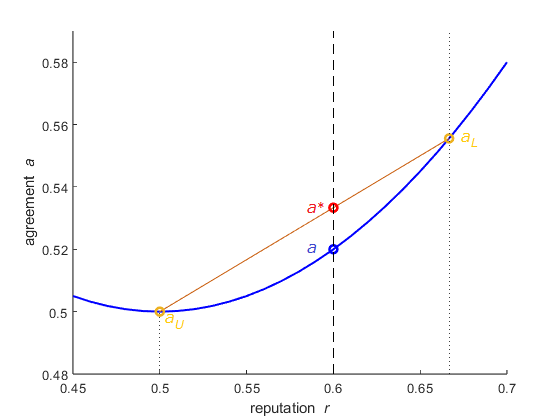}
\caption{Proof sketch that two distinct groups must cause additional agreement.\\
\textmd{(Graphical intuition: Any orange line between two points which are anywhere respectively right and left of the blue dot, will cross the broken black line above said blue dot.)\\
1) The slope of the minimal agreement as a function of reputation $\widehat{a}(r)$ (blue line) is monotonically increasing  ($\widehat{a}(x)=r^2+(1-r)^2$, $\widehat{a}'(x)=4r-2$, $\widehat{a}''(x)=4$).\\
2) The agreement $a$ of a population in which all players have the same reputation (blue circle) is minimal.\\
3) If such a population is split into two groups with different reputations (orange circles), keeping  global $r$ constant (broken black line), the average reputation of one group must decrease ($r_U$) and that of the other must increase ($r_L$).\\
4) The resulting average agreement $a^*$ of the two groups is given by their size $s_U, s_L$ and their particular average agreements $a_U, a_L$. ($a^*=s_Ua_U+s_La_L$). The resulting population state (red circle) lays on the direct line between the group states (orange circles).\\ 
5) The average reputation of the resulting population state remains constant (broken black line), hence the resulting population state is the crossing between orange and broken black line.\\
6) Any agreement of a group ($a_U, a_L$) cannot be lower than the minimal agreement $\widehat{a}$ (blue line) for any average reputation within that group.\\ \\
Therefore, the agreement $a^*$ of the resulting population (red circle) must be higher than the agreement $a$ of the original population (blue circle). 
}}
\label{figSproof}
\end{center}
\end{figure*}

\begin{figure*}[t]
\begin{center}
\includegraphics[width=1\textwidth]{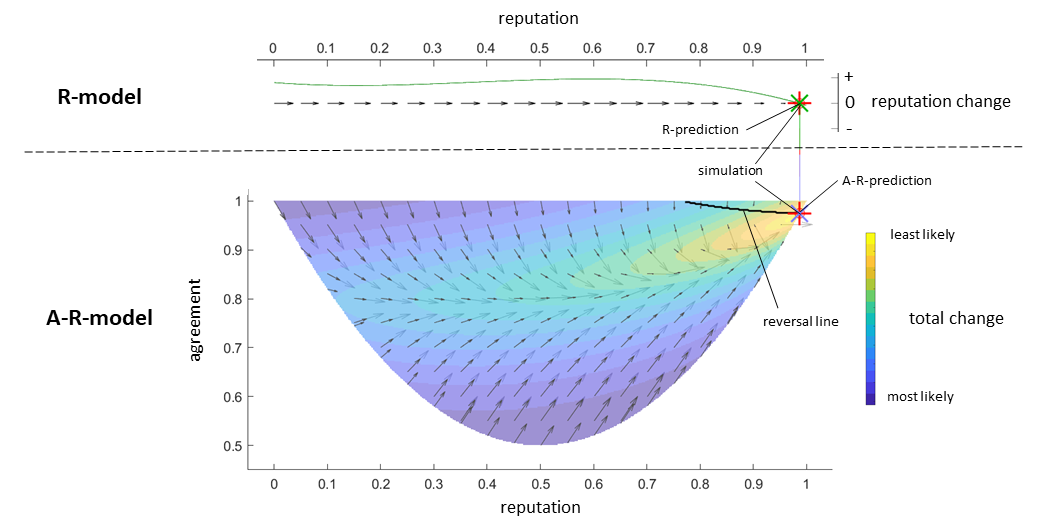}
\caption{A strategy in which reputation can increase if a player agreed less. \textmd{Detailed comparison between models as in Figure 4 in the main text. This strategy, $\alpha=(1,1,0,-1)$, $\beta=(1,0)$, aka "GKGB" was found to be evolutionarily stable (ESS) under private assessment \citep{Okada2020a} while it is not stable under public assessment \citep{Ohtsuki2004a}. (Note, the former study used replicator dynamics between this strategy and AllC and AllD, whereas the latter investigated ESS within different 16 action rules). Our graph shows, that higher agreement sometimes corresponds to decrease in reputation. Above the black line, reputations tend to decrease, while below it they tend to increase (Which is the opposite to the reversal as in Figure 4, which was indicated as a white line). For pubic assessment, we expect lower reputations amongst conditional cooperators with this norm, which are likely to decrease their evolutionary stability. 
Parameters are standard, namely, $\psi=1$, $\epsilon_p=0.1$ and $\mu=0.01$. Bottom graph shows the two-dimensional space of the A-R-model. Dark blue areas indicate likely changes in either agreement or reputation or both, whereas light yellow areas indicate little to no change. Arrows indicate the direction of the change. The precise predictions of the model is indicated by a blue x. Simulations are shown in red, the averages of single runs as red circles and their total average as a red cross. Top graph shows the one-dimensional R-model for comparison. Change in reputation and its direction is depicted on the y-axis in addition to the arrows. The stable point (green diagonal cross) is where the x-axis is crossed with a downward slope.\\
Since A-R-model and R-model converge to the same prediction, results obtained with the R-model \citep{Okada2020a} might also hold for higher observation rates.}}
\label{figS1a}
\end{center}
\end{figure*}

\begin{figure*}[t]
\centering
  \includegraphics[width=0.9\linewidth]{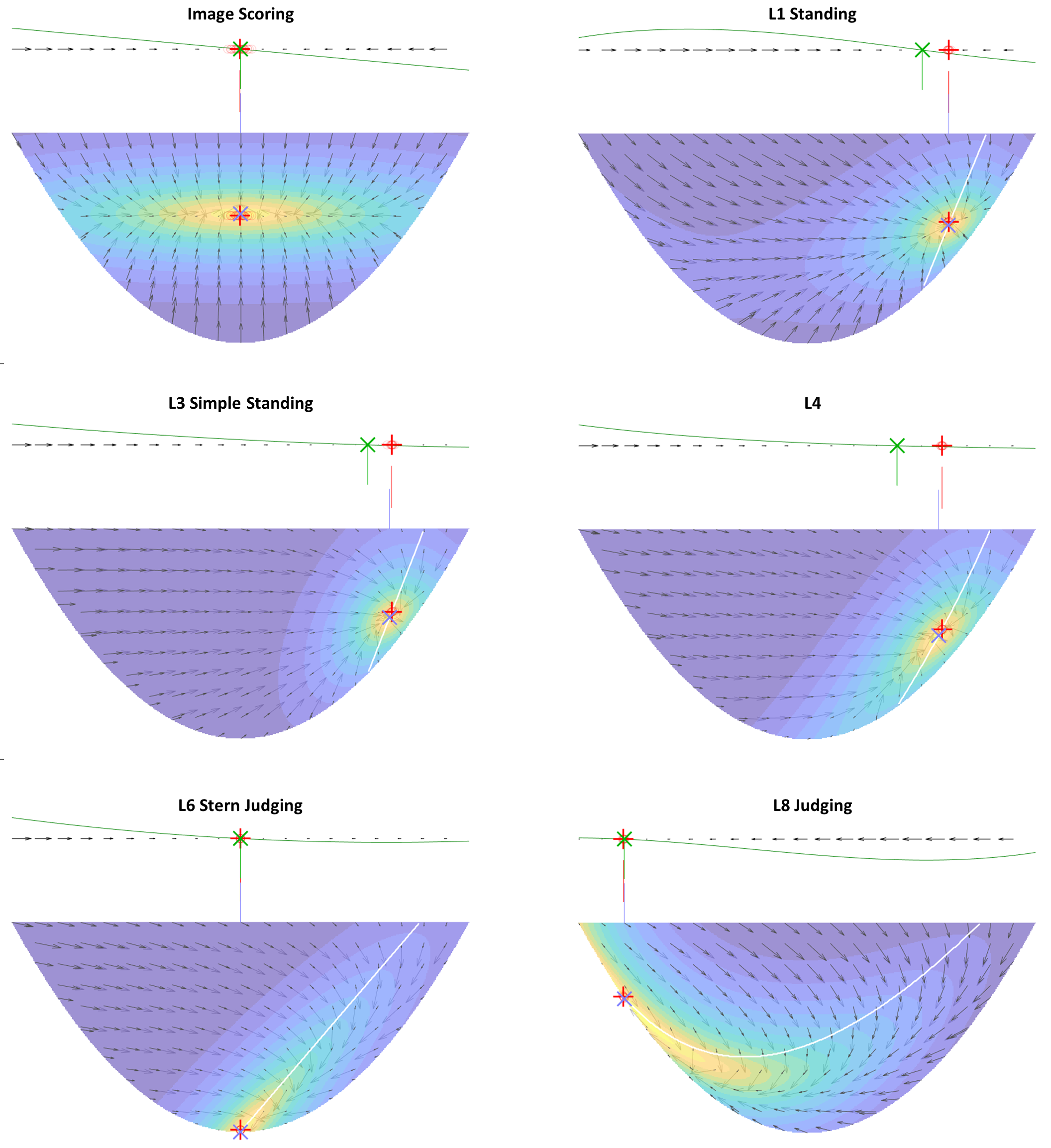}
  \caption{Important strategies. \textmd{Detailed comparison, similar to figure 4 in the main text, between R-model on top and A-R-model below. Shown are the image scoring strategy \citep{Nowak1998} and most leading-eight norms \citep{Ohtsuki2006} (L7 was already depicted in the main text, L2 and L5 cannot be modelled with the current framework, since they require assessment rules that consider the current opinion about the donor, but could be modeled in the future). Parameters are the same as in figure 4: $\psi=1$, $\epsilon_p=0.1$ and $\mu=0.01$.}\\ 
  Image Scoring\textmd{, $\alpha=(1,-1,1,-1)$, $\beta=(1,0)$, is an example of a strategy, where the average reputation is not affected by agreement dynamics. Both R- and A-R-model predict the exact value as the simulations. Image scoring might always behave very similarly under public and under private assessment.}\\
  Standing \textmd{, $\alpha=(1,-1,1,0)$, $\beta=(1,0)$}, Simple Standing\textmd{, $\alpha=(1,-1,1,1)$, $\beta=(1,0)$} and L4 \textmd{, $\alpha=(1,-1,0,1)$, $\beta=(1,0)$, all show similar patterns. The A-R-model is more accurate. And the white reversal line indicates, that higher agreement tends to increase reputations.}\\
  Stern Judging \textmd{, $\alpha=(1,-1,-1,1)$, $\beta=(1,0)$} and Judging \textmd{, $\alpha=(1,-1,-1,0)$, $\beta=(1,0)$, show a slightly different pattern. Higher agreement tends to increase reputation. But, under private assessment, agreement itself tends to fall to a minimal level. The difference between actually reputation under private assessment and reputation under public assessment (presumably where the white reversal line would cross the upper limit of the population state space $a=1$) is especially stark. That corresponds to results that these strategies are no longer stable under private assessment \citep{Hilbe2018}.}
  }
  \label{figS1}
\end{figure*}

\begin{figure*}[t]
\centering
  \includegraphics[width=0.9\linewidth]{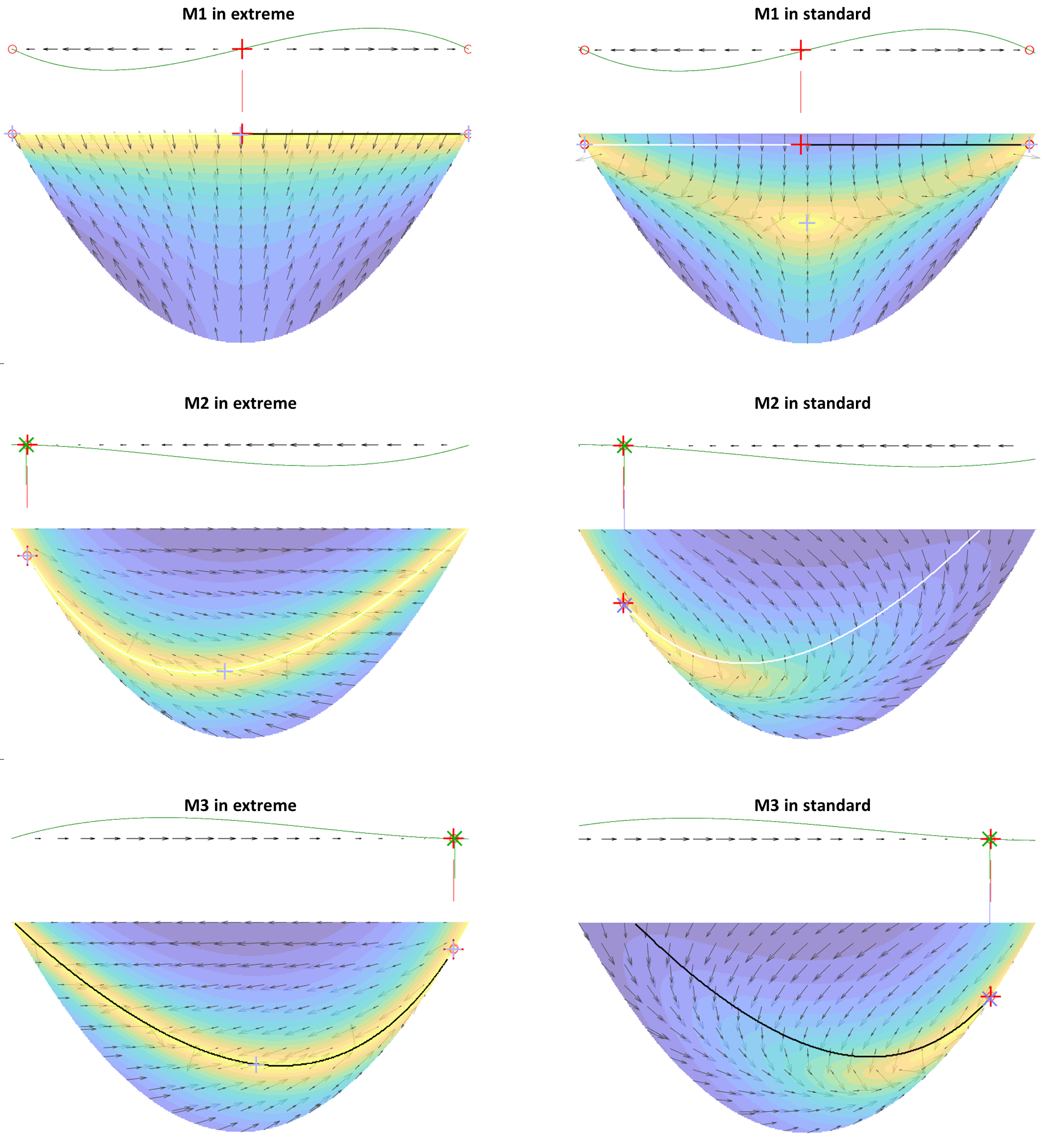}
  \caption{Strategies with multiple equilibria. \textmd{Detailed comparison, similar to Figure 4 in the main text, between R-model on top and A-R-model below. Shown are the three strategies for which the algorithm found multiple equilibria of the A-R-model. Parameters on the left are an 'extreme' condition, $\psi=1$, $\epsilon_p=0$ and $\mu=0.001$, whereas on the right are the 'standard' used in most other figures, $\psi=1$, $\epsilon_p=0.1$ and $\mu=0.01$.} \\
  M1\textmd{, $\alpha=(1,0,0,-1)$, $\beta=(1,0)$, is another strategy found to be evolutionary stable in \cite{Okada2020a}. However, there is no single stable point of the models in most conditions (shown here are only two examples). Neither is there a unique stable region in the simulations. Instead there are two distinct regions, and which one is reached seem to depend on the starting condition (or in the cases of our simulations it is random, since they start in the center, i.e. the same distance to either region). Multiple regions of models and simulations converge. In the models however, there is  a third region, but it is unstable (change is at least numerically close to zero, but all surrounding areas have the expected change that leads away rather than towards this point).}\\
  M2 (aka L8 judging) and M3\textmd{, $\alpha=(1,-1,-1,0)$, $\beta=(1,0)$, and $\alpha=(0,1,1,-1)$, $\beta=(1,0)$, respectively, have multiple equilibria only in a few extreme conditions with a very low perception error $\epsilon$ and very low mutation rate $\mu$. In these conditions, there is a wide band of population states where expected change is at least close enough to zero, so that the numerical algorithm cannot distinguish them easily. For other conditions, such as the standard one, the area of highest stability shrinks and a single stable point can be identified.}
  }
  \label{figS2}
\end{figure*}

\begin{figure*}[t]
\centering
\begin{subfigure}[t]{.48\textwidth}
  \centering
  \includegraphics[width=1\linewidth]{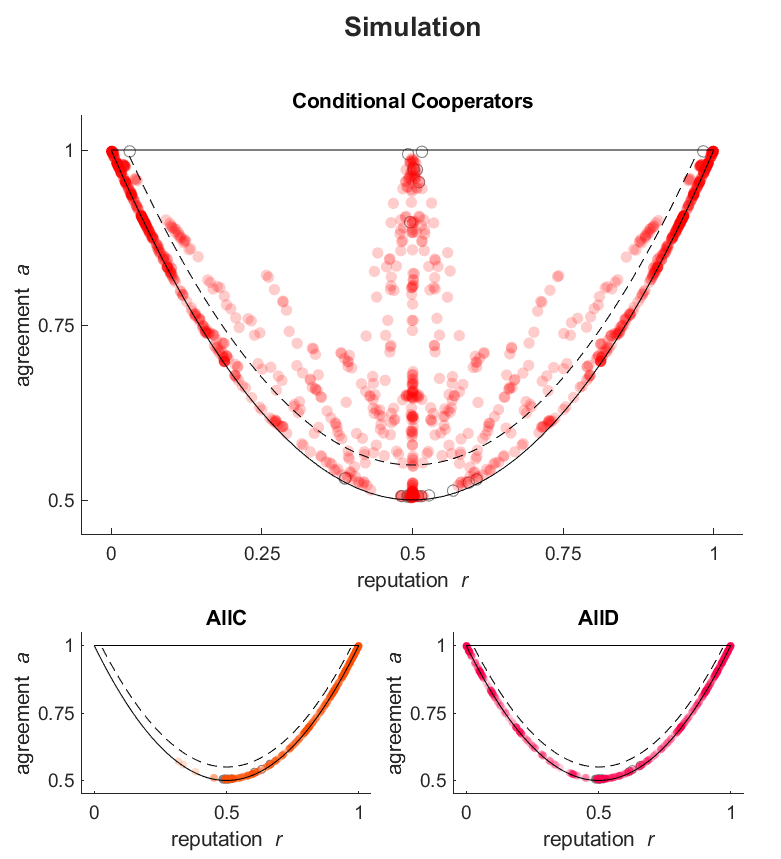}
  \caption{Simulations for all cases. \textmd{Agreement and reputation averaged across runs for homogeneous populations with 100 players. In contrast to figure 2 in the main text, cases where range of averaged reputations between runs exceeded 0.1 are included and shown as grey circles rather than red dots like the other values.}
  }
  \label{fig:sub1}
\end{subfigure}\hfill
\begin{subfigure}[t]{.48\textwidth}
  \centering
  \includegraphics[width=1\linewidth]{fig2P.png}
  \caption{Predictions. \textmd{Unique stable states of the model for infinitely large homogeneous populations. Not shown are cases where multiple equilibria were found which ranged more than 0.01 in total distance in A-R space.}
  }
  \label{fig:sub2}
\end{subfigure}%
\caption{Comparing results of simulations and analytical predictions. \textmd{Values for 171 unique strategies in 15 conditions (namely, $\epsilon \in \{0,0.001,0.01,0.05,0.1,0.2,0.4\}$ and $\mu \in \{0.001,0.01,0.05\}$}).}
\label{figS4}
\end{figure*}

\begin{figure*}
\centering
  \includegraphics[width=1\linewidth]{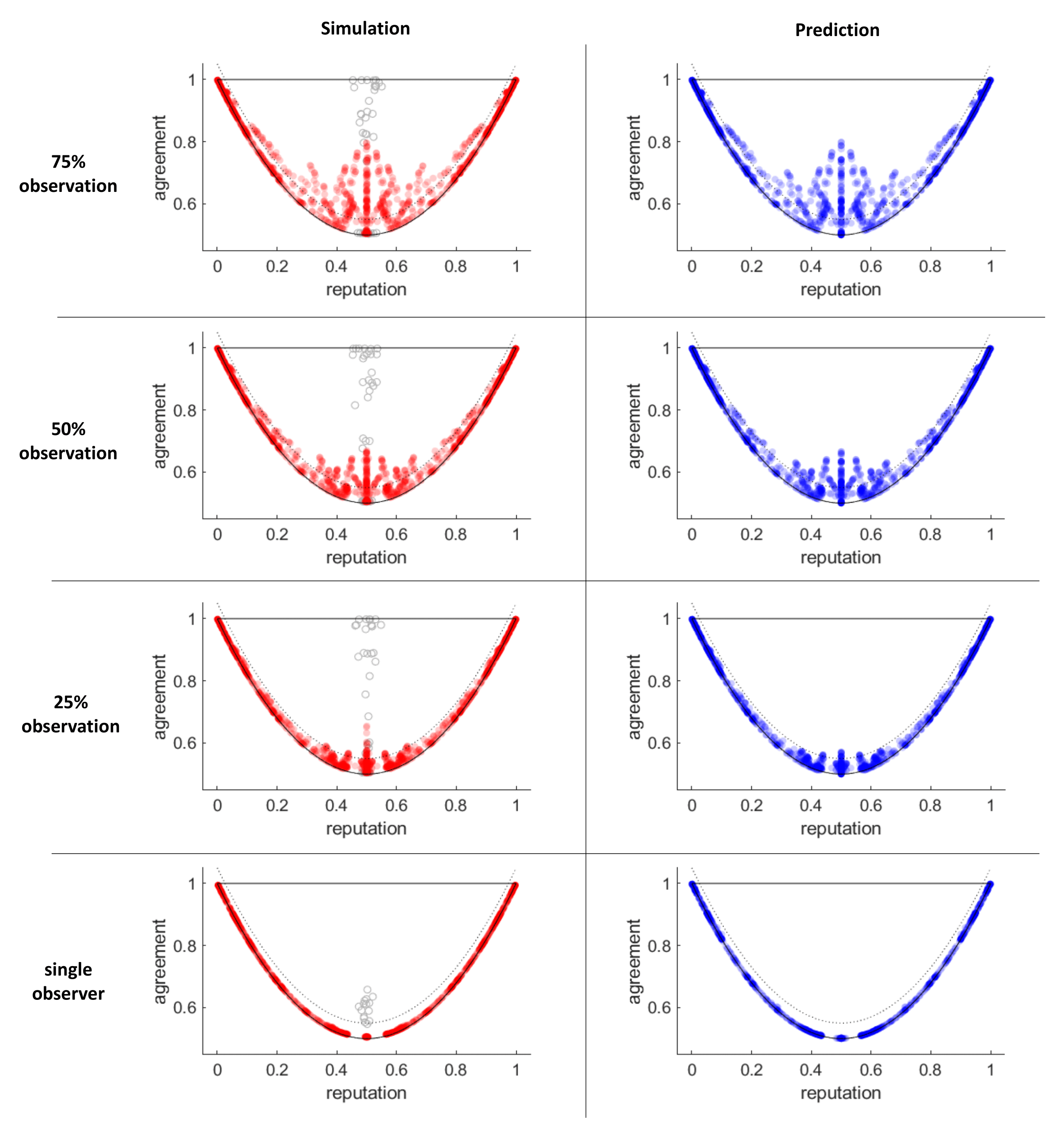}
\caption{Results for Cocos with other observation rates $\psi$. \textmd{On the left are the simulation results, and on the right, the predictions of the A-R-model. Values for all unique 81 conditional cooperators. Values that had been excluded for Figure 3 in the main text are shown as gray circles.}}
\label{figS5}
\end{figure*}

\begin{figure}[t]
\begin{center}
\includegraphics[width=1\linewidth]{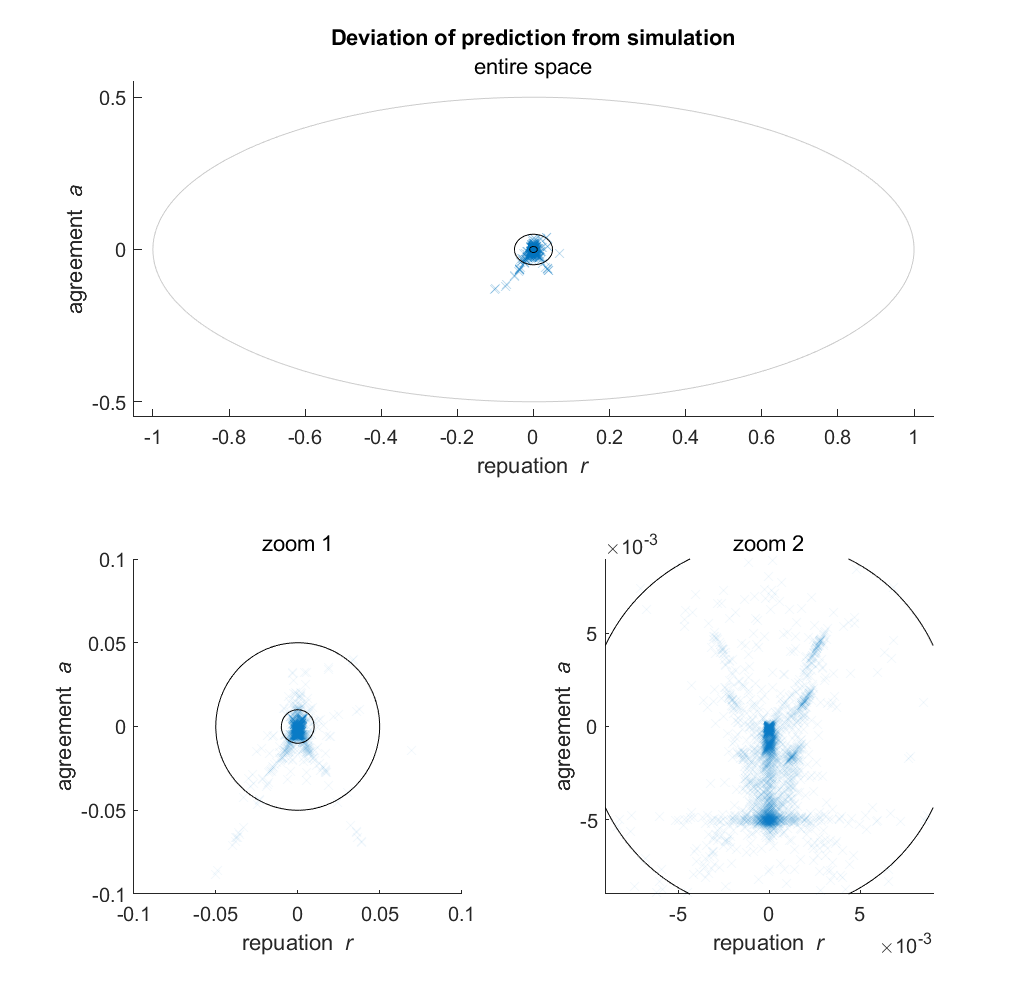}
\caption{Detailed fit for $\psi=1$. \textmd{Shown is both deviation in agreement (y-axis) and deviation in reputation (x-axis). If a case (blue cross) is left and below center, predictions of A-R-model were to small in both agreement and reputation. Large ellipse shows outer boundary of possible prediction deviations, medium circle show total deviation (see equation 3.1) of 0.05 and small circle of 0.01 (for comparison with table 1 in main text). All graphs show the same data, but increasingly zoomed in. (The graph in the lower right corner only shows parts of the most inner circle). As in Figure 2 of the main text, some cases were excluded. \\
The systematic deviations visible on highest zoom could indicate, that the model could still be improved.}
}
\label{figR1S}
\end{center}
\end{figure}

\begin{figure}[t]
\begin{center}
\includegraphics[width=1\linewidth]{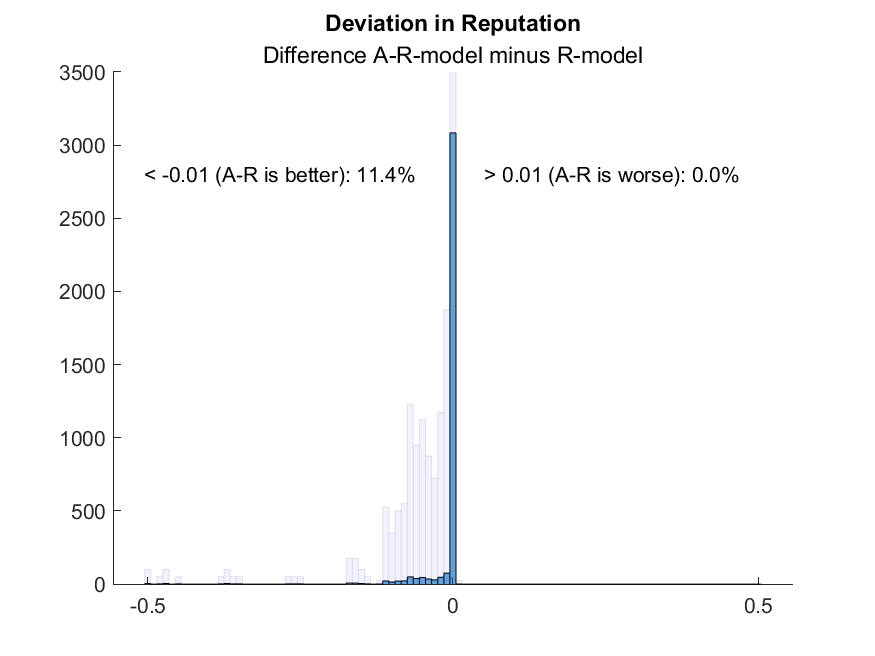}
\caption{Comparison of predictions for reputation between R-model and A-R-model. \textmd{Values are the same as in figure 2 in the main text. Most predictions differ less than 0.01 (center bar of the histogram). But in 11.4\% of predictions, the A-R-model was at least 0.01 more accurate. The light blue 'shadows' of the dark blue bars show the same data but zoomed in, to make the distribution more visible. It shows how some cases are much better predicted by the A-R-model.}}
\label{figBetter}
\end{center}
\end{figure}